\documentclass[prl,twocolumn,aps,superscriptaddress]{revtex4-1}
\usepackage{bm}%
\usepackage{float}
\expandafter\ifx\csname package@font\endcsname\relax\else
 \expandafter\expandafter
 \expandafter\usepackage
 \expandafter\expandafter
 \expandafter{\csname package@font\endcsname}%
\fi

\usepackage{array}
\usepackage{amsmath,amssymb}
\usepackage{MnSymbol}
\usepackage{graphicx}
\usepackage{mathrsfs}
\usepackage{bm}
\usepackage{mathtools}
\usepackage{epstopdf}
\usepackage{hyperref}

\hypersetup{%
   pdfpagemode=None, %FullScreen,
   pdfstartpage=1,
   pdfmenubar=true,
   pdftoolbar=true,
   colorlinks = true,
   linkcolor=blue,
   citecolor=blue,
   urlcolor=blue,
   bookmarksopen=false
 }
 \usepackage{amsmath}
\usepackage{amsfonts}
\usepackage{mathtools}
\usepackage{graphicx}
\usepackage{fixme}
\usepackage{color}
\begin{document}
%\unitlength = 1mm

\title{Attractive and repulsive exciton-polariton interactions mediated  by an electron gas}
\author{Miguel A. Bastarrachea-Magnani}
\affiliation{ 
Department of Physics and Astronomy, Aarhus University, Ny Munkegade, DK-8000 Aarhus C, Denmark. }
\author{Arturo Camacho-Guardian}
\affiliation{ 
Department of Physics and Astronomy, Aarhus University, Ny Munkegade, DK-8000 Aarhus C, Denmark. }
\author{Georg M.\ Bruun}
\affiliation{ 
Department of Physics and Astronomy, Aarhus University, Ny Munkegade, DK-8000 Aarhus C, Denmark. }
\affiliation{Shenzhen Institute for Quantum Science and Engineering and Department of Physics, Southern University of Science and Technology, Shenzhen 518055, China}
\begin{abstract}
Realising strong photon-photon interactions in a solid-state setting is a major goal with  far reaching potential for optoelectronic 
applications. Using Landau's quasiparticle framework combined with a microscopic many-body theory, we explore the interactions between exciton-polaritons and trions in a two-dimensional semiconductor injected with an electron gas inside a microcavity. We show that particle-hole excitations in the electron gas mediate an  attractive interaction between the polaritons, whereas a trion-polariton interaction mediated by the exchange of an electron is either repulsive or attractive depending on the specific polariton branch. These mediated interactions are intrinsic to the quasiparticles
and are also present in the absence of light. Importantly, they can be tuned to be more than an order of magnitude stronger than the direct polariton-polariton interaction in the absence of the electron gas, thereby providing a promising outlook for non-linear optical components. Finally, we compare our theoretical predictions with two recent  experiments. 
\end{abstract}
\maketitle

%%%%%%%%%%%%%%%%%%%%%%%%%%%%%%%%%%%%%%%%%%%%%%%%%%
%%%%%%%%%%%%%%%%%%%%%%%%%%%%%%%%%%%%%%%%%%%%%%%%%%
%%%%%%%%%%%%%%%%%%%%%%%%%%%%%%%%%%%%%%%%%%%%%%%%%%

The realisation of exciton-polaritons in semi-conductor microcavities has opened  a rich setting for hybrid light-matter systems~\cite{Sanvitto2016,Kavokin2017}. Due to their mixed composition, exciton-polaritons provide controllable means to  transfer useful features between light and matter~\cite{Laussy2012,Carusotto2013}, which has led to a range of breakthrough results including the observation of polariton Bose-Einstein condensates (BECs)~\cite{Kasprzak2006,Wouters2007a,Amo2009,Deng2010,Kohnle2011,Kohnle2012}, quantum vortices~\cite{Lagoudakis2008}, and topological states of light~\cite{St-Jean2017,Klembt2018,Ozawa2019,Lubatsch2019}. Monolayer transition metal dichalcogenides (TMDCs) have recently emerged as a particularly promising platform to explore exciton-polaritons, since the excitons are deeply bound and dominate the optical response, and because they offer many spin and valley degrees of freedom~\cite{Wang2018}. A disadvantage of TMDCs however is that the large exciton binding energy and correspondingly small radius makes the direct exciton-exciton interaction weak, which suppresses non-linear optical effects important for optoelectronic devices~\cite{Schaibley2016,Wang2018}. A major challenge is therefore to achieve strong polariton interactions  and several solutions have been suggested including polaritonic Feshbach resonances~\cite{Takemura2014,Navadeh2019},  mediated interactions via polariton BECs~\cite{Camacho-Guardian2020}, and Rydberg excitons~\cite{Walther2018}.

Recently, two experiments have shown that polariton interactions can be enhanced by orders of magnitude by injecting an itinerant electron gas in a monolayer TMDC~\cite{Tan2020,Emmanuele2020}. In these charged systems, the excitons are dressed by the two-dimensional electron gas (2DEG) much like the dressing of impurities in an atomic Fermi gas leading to the formation of Fermi  polarons~\cite{Chevy2006,Massignan2014,Efimkin2017,Chang2018}, and the coupling to light in turn leads to the formation of new quasiparticles coined polaron-polaritons~\cite{Rapaport2001,Qarry2003,Bajoni2006,Sidler2016}. Inspired by these experiments, we combine a microscopic many-body theory with Landau's quasiparticle framework to investigate the interaction between polaritons and trions in a monolayer TMDC injected with a 2DEG inside a microcavity. We show that the interaction between polaritons mediated by particle-hole excitations in the 2DEG is 
attractive, and that the interaction between trions and polaritons mediated by the exchange of an electron can be either attractive or repulsive. Both interactions are an inherent property of the quasiparticles, and they can be tuned to be much stronger than the polariton-polariton interaction in the absence of the 2DEG. We  show that our theory provides a qualitative explanation of the recent experimental results~\cite{Emmanuele2020}. 

%%%%%%%%%%%%%%%
\begin{figure}[!ht]
\begin{center}
\end{center}
\includegraphics[width=0.85\columnwidth]{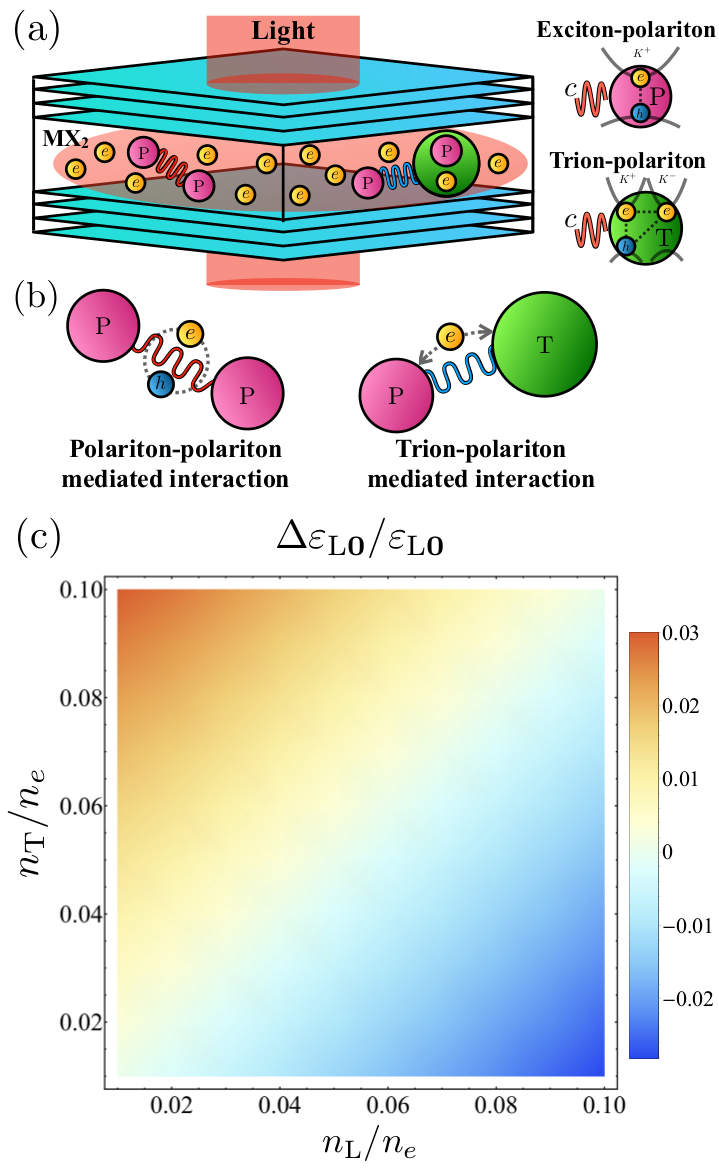}
\caption{(a) Illustration of the system considered. Excitons in a monolayer TMDC are coupled to a cavity light mode forming polaritons (purple balls). The excitons can bind an electron from an itinerant electron gas (yellow balls) to form trion-polaritons (green balls). We show that there are strong polariton-polariton interactions (red wavy line) mediated by particle-hole excitations in the electron gas, as well as polariton-trion interactions (blue wavy line) mediated by the exchange of an electron. (b) The energy shift of a zero momentum L-polariton as a function of its concentration $n_L$ and  the concentration $n_T$ of trions due to these quasiparticle interactions for $\delta/2\Omega=3$.%, and parameters indicated in the main text.
}
\label{fig:1} 
\end{figure}
%%%%%%%%%%%%%%%

%%%%%%%%%%%%%%%
%%%%%%%%%%%%%%%
%%%%%%%%%%%%%%%
\paragraph{System.-} We consider excitons in a TMDC monolayer coupled to a quantum field of light in a microcavity~\cite{Radisavljevic2011,Mak2016,Wang2018}. In addition to the light coupling, the excitons interact with a 2DEG and the Hamiltonian is 
\begin{align} \label{eq:Hamiltonian}
\hat{H}=&\sum_{\mathbf{k}}
\begin{bmatrix}\hat x_{\mathbf{k}}^\dagger&\hat c_{\mathbf{k}}^\dagger\end{bmatrix}
\begin{bmatrix}
\varepsilon_{x\mathbf{k}} & \Omega \\
\Omega & \varepsilon_{c\mathbf{k}}
\end{bmatrix}
\begin{bmatrix}
\hat x_{\mathbf{k}} \\
\hat c_{\mathbf{k}}
\end{bmatrix}
+\sum_{\mathbf{k}}\varepsilon_{e\mathbf{k}}\hat e_{\mathbf{k}}^\dagger\hat e_{\mathbf{k}}
\nonumber \\
&+\frac{1}{2}\sum_{\mathbf{q},\mathbf{k},\mathbf{k}'}V_{\mathbf{q}}\hat{x}_{\mathbf{k}+\mathbf{q}}^{\dagger}\hat{e}_{\mathbf{k}'-\mathbf{q}}^{\dagger}
\hat{e}_{\mathbf{k}'}\hat{x}_{\mathbf{k}},
\end{align}
where $\hat{x}_{\mathbf{k}}^{\dagger}$, $\hat{c}_{\mathbf{k}}^{\dagger}$, and $\hat{e}_{\mathbf{k}}^{\dagger}$ create an exciton, photon, and electron with 2D momentum $\mathbf{k}$. These states have energies $\varepsilon_{x\mathbf{k}}=\mathbf{k}^{2}/2m_{x}$, $\varepsilon_{c\mathbf{k}}=\mathbf{k}^{2}/2m_{c}+\delta$, and $\varepsilon_{e\mathbf{k}}=\mathbf{k}^{2}/2m_{e}$ respectively, where $m_{x}$, $m_{c}$ and $m_{e}$ are the effective masses of the exciton, cavity photon, and electron, and $\delta$ is the detuning between the exciton and photon  at zero-momentum. The energy off-set of the electrons with respect to the excitons will be included in their chemical potential. Due to its
large  binding energy~\cite{Wang2018}, the excitons can be described as point bosons and the spin-valley selection rules in monolayer TMDCs allow us to consider each spin independently~\cite{Cao2012,Yu2015}. The light-matter coupling $\Omega$ is  real, and we use units where $\hbar$, the system area, and $k_B$ are all unity. To model the experiments~\cite{Tan2020,Emmanuele2020}, the exciton-electron interaction $V_{\mathbf{q}}$ is taken to  support a bound state, i.e.\ a trion, with energy $\varepsilon_T^0$. 

We define a $2\times2$ matrix Green's function $\mathcal{G}(\mathbf{k},\tau)=-\langle T_\tau \{\hat\Psi_{\mathbf k}(\tau)\hat\Psi_{\mathbf k}^\dagger(0)\}\rangle$ to describe this strongly interacting hybrid light-matter system, where $\hat\Psi_{\mathbf k}=[\hat x_{\mathbf k},\hat c_{\mathbf k}]^T$ and $T_\tau$ denotes imaginary time ordering. In frequency space,
\begin{align}\label{GreensFn}
\mathcal{G}^{-1}(k)=
\begin{bmatrix}
i\omega_k-\varepsilon_{x\mathbf{k}}- \Sigma(k)& \Omega \\
\Omega & i\omega_k-\varepsilon_{c\mathbf{k}}
\end{bmatrix}
\end{align}
where $k=(\mathbf k,i\omega_k)$ with $i\omega_k$ a bosonic  Matsubara frequency. The self-energy $\Sigma(k)=T\sum_{\mathbf q,i\omega_q}\mathcal{G}_{e}(q)\mathcal{T}(k+q)$ describes the interaction of  excitons with the 2DEG, where $\mathcal{G}_{e}^{-1}(q)=i\omega_q-\xi_{e\mathbf{q}}$ is the electron Green's function with $\xi_{e\mathbf{q}}=\varepsilon_{e\mathbf{q}}-\mu_e$ and $\mu_e$ their chemical potential. The temperature is $T$ and $\mathcal{T}(k)$  the exciton-electron scattering matrix in the ladder approximation including light coupling~\cite{Bastarrachea-Magnani2019,SM}. Performing the frequency sum and  the standard analytic continuation $i\omega_{q}\rightarrow \omega+i0^{+}$ yields
\begin{align} \label{eq:Selfenergy2}
\Sigma(\mathbf k,\omega)= &\sum_{\mathbf q}\left[f(\xi_{e\mathbf{q}})\mathcal{T}(\mathbf{k}+\mathbf{q},\omega+\xi_{e\mathbf{q}})-\frac{f(\xi_{T\mathbf{k}+\mathbf{q}})
\mathcal{Z}_{T\mathbf{k}+\mathbf{q}}}{\omega-\varepsilon_{T\mathbf{k}+\mathbf{q}}+\xi_{e\mathbf{q}}}\right.
 \nonumber \\ 
&\left.+\int_{-\infty}^{\infty}\!\frac{d\omega'}{\pi}\,\frac{f(\omega')\mbox{Im}\mathcal{T}(\mathbf{k}+\mathbf{q},\omega'+i0^{+})}{\omega-\omega'+\xi_{e\mathbf{q}}}\right]
\end{align}
where $f(x)=(\exp x+1)^{-1}$ is the Fermi function. Here $\varepsilon_{T\mathbf{k}}$ is the energy of a trion with momentum $\mathbf{k}$ as determined from the pole of the scattering matrix $\mathcal{T}(\mathbf{k})$, which differs from $\varepsilon_{T\mathbf{k}}^0=\varepsilon_T^0+k^2/2m_T$ where $m_T=m_x+m_e$ is the trion mass, 
due to the presence of the 2DEG and the light coupling, $\mathcal{Z}_{T\mathbf{k}}$ is its residue in the exciton channel, and $\xi_{T\mathbf{k}}=\varepsilon_{T\mathbf{k}}-\mu_T$ with $\mu_T$ the trion chemical potential. 

%%%%%%%%%%%%%%%
%%%%%%%%%%%%%%%
\begin{figure}[!ht]
\begin{center}
\end{center}
\includegraphics[width=0.85\columnwidth]{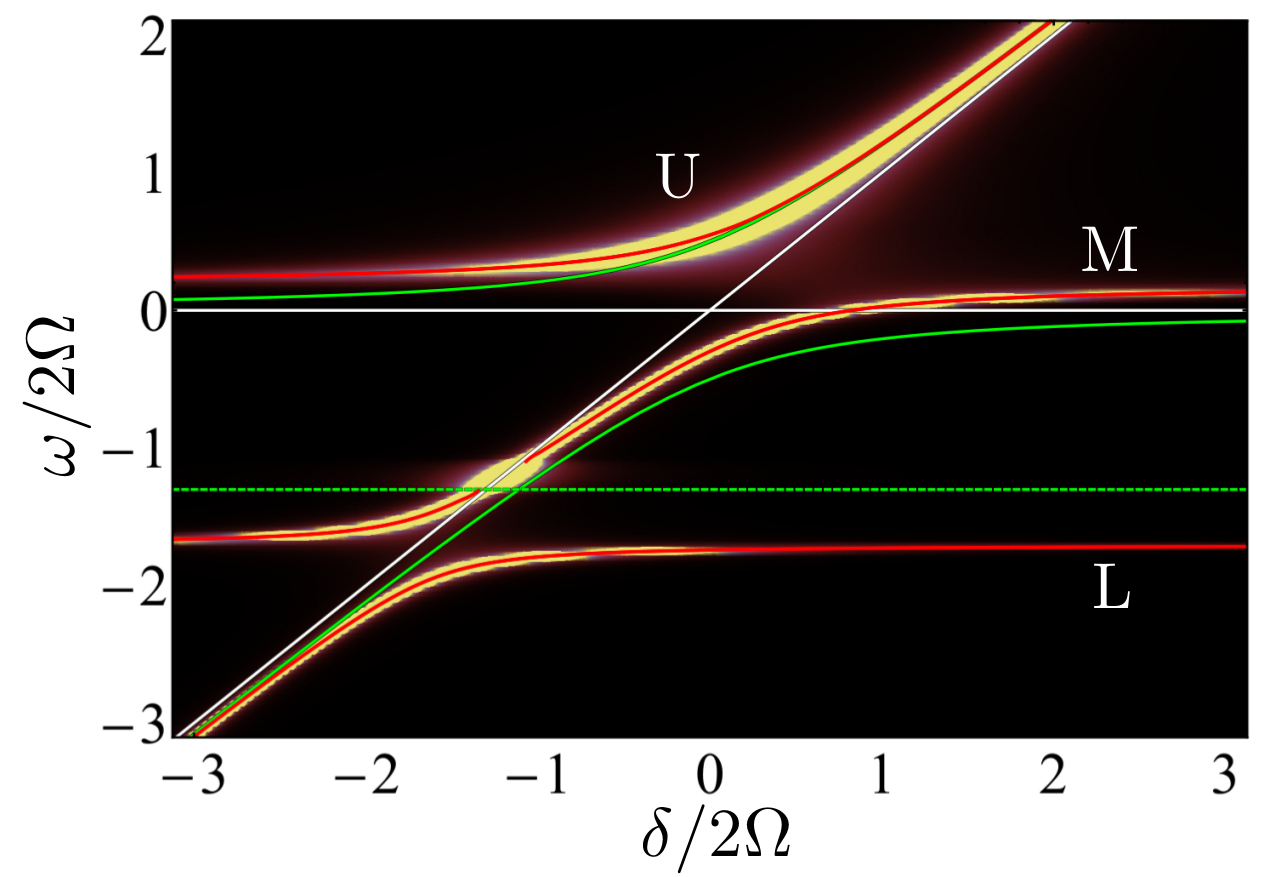}
\caption{Zero momentum spectral function of the cavity photons $A_{cc}(0,\omega)$ as a function of  energy and detuning. The white lines give the uncoupled photon and exciton energies, the green lines indicates the upper and lower polariton energies in the absence of the 2DEG, and the red lines give the polaron-polariton energies in the presence of the 2DEG. The horizontal green dashed line marks the energy of the trion plus an electron from the 2DEG.}
\label{fig:2} 
\end{figure}
%%%%%%%%%%%%%%%
%%%%%%%%%%%%%%%

%%%%%%%%%%%%%%%
%%%%%%%%%%%%%%%
%%%%%%%%%%%%%%%
\paragraph{Polaron-polaritons.-} Before we investigate the interactions between the quasiparticles of the system, we briefly discuss their single particle properties. In Fig.~\ref{fig:2}, 
we plot the zero temperature 
photon spectral function $A_{cc}(\mathbf k=0,\omega)=-2\text{Im}G_{cc}(\mathbf k,\omega)$ obtained by inverting Eq.~\eqref{GreensFn} as a function of the detuning $\delta$. We have assumed a single 
quasiparticle so that $f(\omega')=f(\xi_{T\mathbf{k}+\mathbf{q}})=0$ in Eq.~\eqref{eq:Selfenergy2}. Here, and in the rest of the manuscript we use experimentally realistic values $m_{x}=2m_{e}$, $m_{c}=10^{-5}m_{e}$, $\Omega=8\mbox{meV}$, 2DEG density $n_{e}=8\mbox{x10}^{11}\mbox{cm}^{-2}$ corresponding to $\mu_e/2\Omega=0.23$, and trion energy $\varepsilon_{T}^0=-25\mbox{meV}$ ($\mu_e/\varepsilon_{T}^{0}=0.15$)~\cite{Sidler2016,Tan2020,Mak2012,Wang2015}. For the calculations, we add a small imaginary part $\eta/2\Omega=0.01$ to the frequency. The lines in Fig.~\ref{fig:2} give the energy $\varepsilon_{i\mathbf{k}=0}$ of a quasiparticle  branch $i$, which can be obtained by solving self-consistently $\varepsilon_{\mathbf{k}}=\left(2\varepsilon_{c\mathbf{k}}-\delta_{\mathbf{k}}\pm\sqrt{\delta_{\mathbf{k}}^{2}+4\Omega^{2}}\right)/2$  with $\delta_{\mathbf{k}}=\varepsilon_{c\mathbf{k}}-\varepsilon_{x\mathbf{k}}-\Sigma(\mathbf k,\varepsilon_{\mathbf{k}})$~\cite{SM}. 
They are shifted away from the bare upper and lower polariton energies in the absence of the 2DEG ($\Sigma(k)=0$) due to electron-exciton interactions. When $|\delta|\gg2\Omega$, the excitons and the photons decouple and the three quasiparticle branches correspond to the photon, an attractive polaron with an energy below the trion energy minus that of electron from the 2DEG (green dashed line at $\varepsilon=\varepsilon_{T}^0+(m_{e}/m_{x}+m_{e}/m_{T})\mu_e$, 
and a repulsive polaron with  an energy above the bare exciton energy in direct analogy with polarons in atomic Fermi gases~\cite{Massignan2014,Schmidt2012}. For smaller detuning $|\delta|\lesssim 2\Omega$, the polarons and photons hybridise giving rise to three quasiparticle branches, which we refer to as the lower (L), middle (M) and upper (U) polaritons. The M-polariton becomes ill-defined when  $\delta/2\Omega\simeq-1.2$ where its energy plus that of an electron from the 2DEG matches the trion state. These quasiparticle  have recently been observed experimentally~\cite{Sidler2016,Tan2020}. 

%%%%%%%%%%%%%%%
%%%%%%%%%%%%%%%
%%%%%%%%%%%%%%%
\paragraph{Quasiparticle interactions.-} An intrinsic property of quasiparticles is that they interact via excitations in the surrounding medium, which in the present case is the 2DEG. As first pointed out by Landau, it follows that the energy of a quasiparticle in branch $i$ and with momentum ${\mathbf k}$ can be written as~\cite{Landau1957,Baym1991} 
\begin{align}
\label{LandauTP}
\varepsilon_{i\mathbf k}=\varepsilon_{i\mathbf k}^{0}+\sum_{j\mathbf k'}\mathsf{f}_{i\mathbf k,j\mathbf k'}\delta n_{j\mathbf k'}+\ldots,
\end{align}
where $\varepsilon_{i\mathbf k}^{0}$ is the energy for vanishing quasiparticle concentrations, $\mathsf f_{i\mathbf{k},j\mathbf{k}'}$ is the interaction between quasiparticles in branches $i$ and $j$ with momenta $\mathbf{k}$ and $\mathbf{k}'$, and $\delta n_{i\mathbf k}$ is the quasiparticle distribution function. 

To connect this general expression with our microscopic theory, we observe that the self-energy in Eq.\ \eqref{eq:Selfenergy2} depends explicitly on the trion distribution function $f(\xi_{T\mathbf{k}})$  and on the polariton distribution functions through the many-body scattering matrix $\mathcal{T}(\mathbf{k},\omega)$~\cite{SM}. The quasiparticle energies therefore depend on these densities, which in terms of Eq.~\eqref{LandauTP} corresponds to polariton-polariton and polariton-trion interactions mediated by the 2DEG. 
Focusing for concreteness on a L-polariton with momentum $\mathbf k$, its mediated interaction  with another polariton or trion with momentum $\mathbf k'$ can be calculated microscopically as~\cite{SM}
\begin{align} \label{MicroscopicInt}
\mathsf f_{L\mathbf{k},j\mathbf{k}'}
=\mathcal Z_{L\mathbf{k}}\mathcal{C}_{\mathbf{k}}^{2}\frac{\delta\Sigma(\mathbf{k},\varepsilon_{L\mathbf{k}})}{ \delta n_{j{\mathbf{k}'}}},
\end{align}
where $\mathcal Z_{L\mathbf{k}}^{-1}=1-\mathcal{C}^{2}_{\mathbf{k}}\partial_\omega\Sigma(\mathbf{k},\omega)|_{\varepsilon_{L\mathbf{k}}}$ is the residue of the L-polariton. Compared to the usual microscopic expression for the quasiparticle interaction from the  self-energy~\cite{Giuliani2005,Camacho2018}, Eq.~\eqref{MicroscopicInt} has an additional Hopfield factor $\mathcal C^{2}_{\mathbf k}=(1+\delta_{\mathbf{k}}/\sqrt{\delta_{\mathbf{k}}^2+4\Omega^{2}})/2$ reflecting that it is only the exciton component of the L-polariton that interacts with the 2DEG. We now analyse the polariton-polariton and the polariton-trion interactions in detail.  
%%%%%%%%%%%%%%%
%%%%%%%%%%%%%%%
%%%%%%%%%%%%%%%

\begin{figure}[!ht]
\begin{center}
\end{center}
\includegraphics[width=0.85\columnwidth]{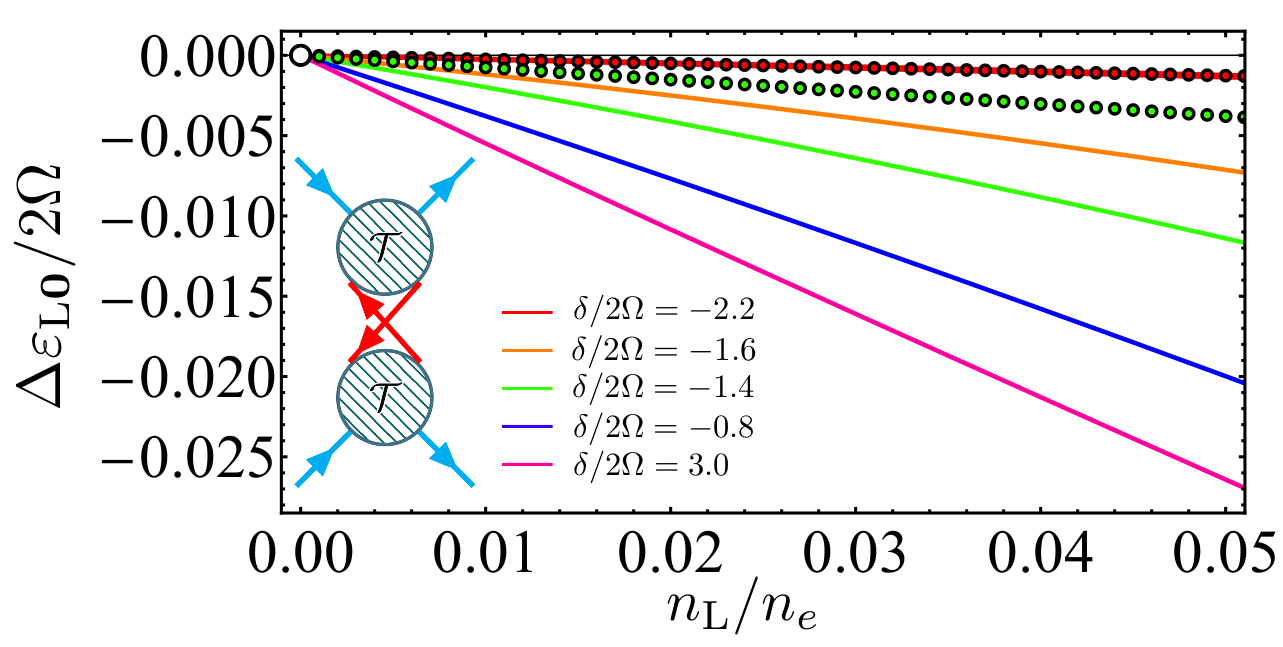}
\caption{The energy shift of a zero momentum L-polariton as a function of its density for various detunings. The dotted lines are the perturbative result obtained from Eqs.~\eqref{LandauTP}-\eqref{RKKY} for $\delta/2\Omega=-2.2$ (red) and $-1.4$ (green). The Feynman diagram for the attractive polariton-polariton interaction mediated by electron particle-hole 
excitations giving rise to these energy shifts is shown in the inset. Blue/red lines are exciton/electron propagators and the grey circles the exciton-electron scattering matrix~\cite{SM}.}
\label{fig:3} 
\end{figure}

\begin{figure}[!ht]
\begin{center}
\end{center}
\includegraphics[width=0.75\columnwidth]{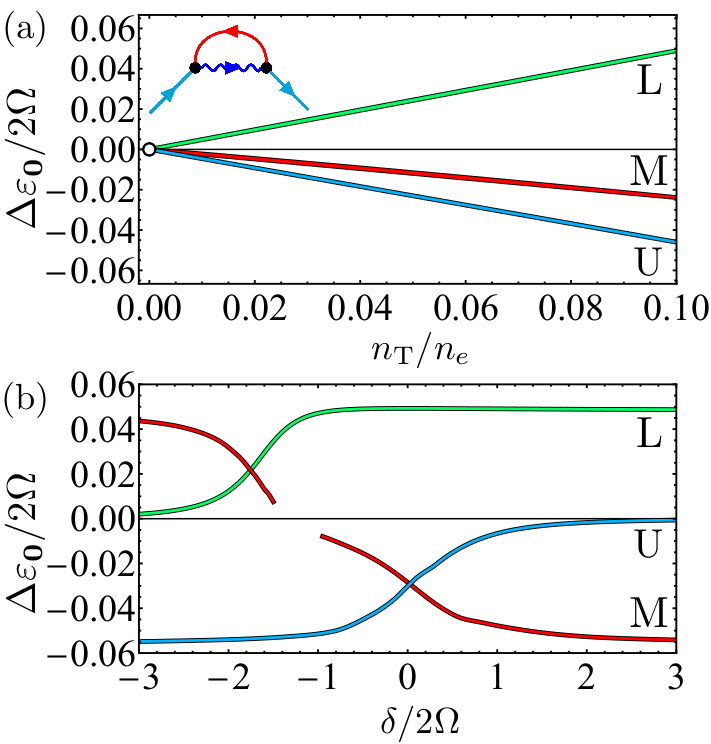}
\caption{(a) Energy shift of the three polariton branches for zero momentum  as a function of the trion density for $\delta/2\Omega=-0.5$.  The inset shows the Feynman diagram for the  trion-polariton interaction giving rise to these energy shifts. The blue curvy line is the trion propagator~\cite{SM}.
(b) Energy shift as a function of cavity detuning for $n_T/n_e=0.1$.}
\label{fig:4} 
\end{figure}

\paragraph{Polariton-polariton interaction.-}
We first examine the interaction between L-polaritons mediated by the 2DEG. In Fig.~\ref{fig:3}, the energy shift $\Delta\varepsilon_{L\mathbf k}$ of the $\mathbf k=0$ L-polariton is plotted as a function of its concentration $n_{L}$ for various detunings $\delta$. It is calculated by evaluating Eqs.~\eqref{GreensFn}-\eqref{eq:Selfenergy2} numerically with a varying polariton density given by $n_{L}=\sum_\mathbf{k}f(\varepsilon_{L\mathbf{k}}-\mu_L)$, where we have introduced the  chemical potential  $\mu_{L}$ of the L-polaritons. The temperature is $T=0.1\mu_{e}$ so that the L-polaritons 
remain uncondensed. The energy is seen to decrease with the $n_{L}$ showing that the interaction between the  L-polaritons is \emph{attractive}. Taking the functional derivative in Eq.~\eqref{MicroscopicInt} with $\delta n_{j{\mathbf{k}'}}=\delta n_{L{\mathbf{k}'}}=f(\varepsilon_{L\mathbf{k}'}-\mu_L)$ of the self-energy give by Eq.~\eqref{eq:Selfenergy2} yields the diagram shown in the inset of 
Fig.~\ref{fig:4}~\cite{SM}. It describes a quasiparticle interaction mediated by a particle-hole  excitations in the 2DEG, which is inherently  attractive since it corresponds to one quasiparticle creating a density modulation that attracts the other quasiparticle. 
%%%%%%%%%%%%%%%
%%%%%%%%%%%%%%%
%%%%%%%%%%%%%%%
%%%%%%%%%%%%%%%

When $\delta/2\Omega\ll -1$, the energy of the L-polariton is far detuned from the trion energy, see Fig.~\ref{fig:2}, and the exciton-electron scattering matrix $\mathcal T$ depends only weakly on the energy. Equation~\eqref{MicroscopicInt} then yields~\cite{SM}
\begin{gather}
\label{RKKY}
\mathsf f_{L\mathbf k,L\mathbf k'}=\mathcal C_{\mathbf k}^2\mathcal C_{\mathbf k'}^2\mathcal{T}_{n_{L}=0}^{2}(\mathbf 0,\varepsilon_{L0})\chi(\mathbf{k}-\mathbf{k}',\varepsilon_{L\mathbf{k}}-\varepsilon_{L\mathbf{k}'})
\end{gather}
for the interaction between two L-polaritons with momenta $\mathbf{k}$ and $\mathbf{k}'$, where $\chi(k)$ is the 2D Lindhard function~\cite{Giuliani2005}. Equation \eqref{RKKY} is the usual Ruderman-Kittel-Kasuya-Yosida (RKKY) interaction mediated by a degenerate electron gas except for the Hopfield coefficients, which again reflect that it is only the exciton component of the L-polaritons that interacts with the electrons. Figure \ref{fig:4} shows that the weak coupling result obtained by using Eq.~\eqref{RKKY} 
in  Eq.~\eqref{LandauTP}  agree with the energy shift  obtained from the full self-energy for $\delta/2\Omega=-2.2$. The agreement breaks down for   $\delta/2\Omega=-1.4$, where the energy of the L-polariton approaches that of the trion, see Fig.~\ref{fig:2}, and the scattering matrix depends significantly on the energy leading to a strong mediated interaction. The interaction is further strengthened by the fact that the excitonic component of the L-polariton increases with the detuning as determined by the Hopfield coefficient ${\mathcal C}_\mathbf{k}$. Similar calculations show that the interaction between M- and U-polaritons mediated by electron-hole excitations in the 2DEG is also attractive as expected~\cite{Yu2012}.

%%%%%%%%%%%%%%%
%%%%%%%%%%%%%%%
%%%%%%%%%%%%%%%
\paragraph{Polariton-trion interaction.-} 
Consider now the interaction between polaritons and trions. In Fig.~\ref{fig:4}(a), the zero temperature energy shift of all three polariton branches due to a non-zero trion density $n_T=\sum_\mathbf{k}f(\xi_{T\mathbf k})$ is plotted for $\delta/2\Omega=-0.5$. We see that the energy of the L-polariton increases with $n_T$ corresponding to a \emph{repulsive} interaction with the trions, whereas the energy of the M- and U-polaritons decrease corresponding to an \emph{attractive} interaction.

To understand this,  focus on the L-polaritons first. Their interaction with the trions can be obtained from Eqs.~\eqref{eq:Selfenergy2} and~\eqref{MicroscopicInt} 
with $\delta n_{j{\mathbf{k}'}}=\delta n_{T{\mathbf{k}'}}=f(\xi_{T\mathbf{k}})$ giving
 \begin{align}\label{TrionPolaritonInt}
\mathsf f_{L\mathbf k,T\mathbf k'}=-\mathcal Z_{L\mathbf{k}}\mathcal{C}_{\mathbf{k}}^{2}\frac{\mathcal{Z}_{T\mathbf{k}'}}{\varepsilon_{L\mathbf k}-\varepsilon_{T\mathbf{k}'}+
\xi_{e\mathbf{k}'-\mathbf{k}}}.
\end{align}
Physically, this interaction  originates from the  coupling of a polariton with momentum $\mathbf k$ and an electron with momentum $\mathbf k'-\mathbf k$ to a  trion with momentum $\mathbf k'$. The resulting second order energy shift of the L-polariton is negative since its energy plus that of the electron is always lower than the trion energy,  see Fig.~\ref{fig:2}. However, a non-zero trion density Fermi blocks  this coupling thereby reducing the negative energy shift, which corresponds to a \emph{repulsive} interaction between a trion and a L-polariton mediated by the exchange of an electron. The corresponding Feynman diagram for this mediated interaction is shown in the inset of Fig.~\ref{fig:3}. The same reasoning also explains why the interaction between the M- and U-polaritons and the trions is attractive. Indeed, since their energies plus an electron from the 2DEG is above that of the trion for the parameters chosen, see Fig.~\ref{fig:2}, 
 the denominator in Eq.~\eqref{TrionPolaritonInt} is positive and the interaction flips sign compared to the case of L-polaritons~\footnote{The expression for the mediated interaction between trions and M-and U-polaritons is identical to Eq.~\eqref{TrionPolaritonInt} except for a different Hopfield coefficient.}
%~\cite{footnote}. 

Figure \ref{fig:4} (b) shows the energy shift of the polaritons caused by a trion density $n_{T}/n_{e}=0.1$ as a function of the detuning $\delta$. Again, the energy shift of the L-polariton is always positive since its energy plus that of an electron from the 2DEG is below that of the trion state for all $\delta$. The energy shift of the M-polariton shifts from positive to negative around $\delta/2\Omega\sim-1.5$ reflecting that its energy plus that of an electron crosses the trion energy from below, as can be seen in Fig.~\ref{fig:2}. In contrast, the energy shift of the U-polariton is always negative as its energy is above that of the trion for all values of detuning. 
  
The main results of this letter are summarized in Fig.~\ref{fig:1}(c), where we plot the energy shift of the L-polariton as a function of its density $n_L$ and of the trion density $n_T$ for 
$\delta/2\Omega=3$ and $T=0.1\mu_e$. It is calculated numerically from Eqs.~\eqref{GreensFn}-\eqref{eq:Selfenergy2} by varying both chemical potentials $\mu_L$ and $\mu_T$. Figure \ref{fig:1}(c) clearly shows how the energy of the L-polariton increases/decreases with increasing trion/polariton density reflecting the underlying repulsive/attractive quasiparticle interactions. 

%%%%%%%%%%%%%%%
%%%%%%%%%%%%%%%
%%%%%%%%%%%%%%%
%%%%%%%%%%%%%%%

\paragraph{Discussion and outlook.-} 
We can define an effective strength $g_{LL}$ of the mediated interaction between L-polaritons by writing $\Delta\varepsilon_L=g_{LL}n_L$. Note that $g_{LL}$ 
includes strong correlations
 despite its definition from a  mean-field type expression.  From Fig.~\ref{fig:3}, we extract $g_{LL}=-0.4\mu$eV$\mu$m$^2$ for $\delta/2\Omega=-1.4$ and 
$g_{LL}=-1.0\mu$eV$\mu$m$^2$ for $\delta/2\Omega=3.0$. Since polaritons are mainly excitons for $\delta/2\Omega=3.0$, we can compare the latter with the 
experimental values $g_{xx}\simeq 0.05$
for the direct exciton-exciton interaction~\cite{Barachati2018,Tan2020}. This shows that the 2DEG amplifies the interaction by more than an order of magnitude. 
We can also define an effective  interaction between trions and L-polaritons by writing $\Delta\varepsilon_L=g_{LT}n_T$ and 
from Fig.~\ref{fig:4}, we extract  $g_{LT}=0.5\mu$eV$\mu$m$^2$ for $\delta/2\Omega=-0.5$. In Ref.~\cite{Tan2020}, 
 a time-dependent energy shift of strongly damped polaritons was observed, which was attributed to a repulsive interaction $g=0.5\mu$eV$\mu$m$^2$.
This is 50 times larger than the observed polariton-polariton interaction strength   in the absence of the 2DEG, and  was attributed to non-equilibrium effects.

In Ref.~\cite{Emmanuele2020}, the energy of the L- and the M-polariton%~\cite{footnote2} 
was observed to increase and decrease respectively with increasing trion density, which is precisely what our theory predicts, 
see Fig.~\ref{fig:4}~\footnote{The M-polariton was not distinguished from the trion for the specific experimental situation with a low electron density.}. 
We moreover obtain an energy shift of $\Delta\varepsilon_{M\mathbf 0}\simeq -0.2\mbox{meV}$ for $n_T/n_e=$0.05, $\varepsilon_{T}^{0}=-30\mbox{meV}$, and 
%$\delta/2\Omega=-1.6$
$\delta=-15.4\mbox{meV}$, which is of the same order as reported experimentally. This suggests that the observed large  energy shifts
 are due to the strong mediated interactions between the quasiparticles. Note that these interactions are always present also in 
 the absence of light where the polaritons become excitons. In Refs.~\cite{Emmanuele2020,Shahnazaryan2020arXiv}, the energy shifts in the low concentration limit were on the other hand attributed to a reduction of the Rabi coupling between light and trions. It would be very interesting to investigate these energy shifts further. In particular, a pump-probe experiment where both the pump and the probe beams selectively populate the trion or polariton branches could unravel the mechanism behind the  quasiparticle interactions.
 This would constitute an important step towards realising and controlling 
strong polariton interactions  with far reaching perspectives for optoelectronic applications.   

%%%%%%%%%%%%%%%
%%%%%%%%%%%%%%%
%%%%%%%%%%%%%%%
\section*{Acknowledgements}
We acknowledge financial support from the Villum Foundation and the Independent Research Fund Denmark - Natural Sciences via Grant No.\ DFF - 8021- 00233B. We thank J. Thomsen for valuable input and insight at the early stage of the project. Valuable discussions with A. Imamoglu and O. Cotlet are also appreciated. M. A. B. M. is also grateful to N. Ram\'irez-Cruz for her valuable insights. 

\bibliography{FB_Notes}

%%%%%%%%%%%%%%%%%%%%%%%%%%%%%%%%%%%%%%%%%%%%%%%%%%

\widetext
\clearpage
\begin{center}
\textbf{\large Supplemental Materials: Attractive and repulsive exciton-polariton interactions mediated  by an electron gas}
\end{center}
\setcounter{equation}{0}
\setcounter{figure}{0}
\setcounter{table}{0}
\setcounter{page}{1}
\makeatletter
\renewcommand{\theequation}{S\arabic{equation}}
\renewcommand{\thefigure}{S\arabic{figure}}
\renewcommand{\bibnumfmt}[1]{[S#1]}
\renewcommand{\citenumfont}[1]{S#1}

%%%%%%%%%%%%%%%%%%%%%%%%%%%%%%%%%%%%%%%%%%%%%%%%%%

\section{Scattering matrix and trion energy}
\label{app:1}

The self-energy describing the electron-exciton interaction from the main text is given by
\begin{align} \label{Selfenergy}
\Sigma(k)=T\sum_{q}\mathcal{G}_{e}(q)\mathcal{T}(k+q),
\end{align}
where $q=(\mathbf{q},i\omega_{\nu})$, $\omega_{q}$ is a fermionic Matsubara frequency, $\mathcal{G}_{e}^{-1}(q)=i\omega_q-\xi_{e\mathbf{q}}$ is the electron Green's function with $\xi_{e\mathbf{q}}=\varepsilon_{e\mathbf{q}}-\mu_e$, $\mu_e$ the electron chemical potential, and $T$ the temperature. The exciton-electron scattering matrix 
\begin{gather} 
\mathcal{T}(k)=\left[\text{Re}\Pi_v(\varepsilon_{T}^{0})-\Pi(k)\right]^{-1}, 
\label{eq:Tmatrix}
\end{gather} 
is related to the electron-exciton pair propagator 
\begin{gather} \label{Pairpropagtor}
\Pi(q)=-T\sum_{q}\mathcal{G}^{(0)}_{x}(k+q)\mathcal{G}_{e}(-q),
\end{gather}
by means of a Dyson equation $\mathcal{T}(k)=\mathcal{T}_{0}+\mathcal{T}_{0}\Pi(k)\mathcal{T}(k)$ resulting from the ladder approximation~\cite{Fetter1971}. Here, $\mathcal{G}_{x}^{0}(k)=\sum_{i}(\mathcal{X}_{i\mathbf{k}}^{0})^{2}/(i\omega_{k}-\xi^{0}_{i\mathbf{k}})$
is the exciton propagator in absence of interactions, but including the light-matter coupling, with $i=L,U$, $\xi^{0}_{i\mathbf k}=\varepsilon^{0}_{i\mathbf k}-\mu_P$, $\varepsilon^{0}_{L,U\mathbf{k}}=\left(2\varepsilon_{c\mathbf{k}}-\delta_{\mathbf{k}}^{0}\pm\sqrt{(\delta_{\mathbf{k}}^{0})^{2}+4\Omega^{2}}\right)/2$, $\delta_{\mathbf{k}}^{0}=\varepsilon_{c\mathbf{k}}-\varepsilon_{x\mathbf{k}}$, and $(\mathcal{X}_{L\mathbf{k}}^{0})^{2}=(\mathcal{C}_{\mathbf{k}}^{0})^{2}=(1+\delta^{0}_{\mathbf{k}}/\sqrt{(\delta_{\mathbf{k}}^{0})^{2}+4\Omega^{2}})/2$ and $(\mathcal{X}_{U\mathbf{k}}^{0})^{2}=(\mathcal{S}_{\mathbf{k}}^{0})^{2}=(1-\delta^{0}_{\mathbf{k}}/\sqrt{(\delta_{\mathbf{k}}^{0})^{2}+4\Omega^{2}})/2$ are the standard Hopfield coefficients in absence of interactions~\cite{Hopfield1958}. 

We eliminate the UV divergence in the scattering matrix by renormalizing the propagator using the energy $\varepsilon_{T}^{0}$ of 
  the bound state, i.e.\  the trion, which comes from the solution of the scattering problem of a exciton-electron pair in the vacuum
\begin{gather}
\mbox{Re}\Pi_{v}(\varepsilon_{T}^{0})=-\frac{\mu}{2\pi}\log\left(\frac{\Lambda^{2}}{2\mu|\varepsilon_{T}|}\right).
\end{gather}
Here $\mu^{-1}=m_{e}^{-1}+m_{x}^{-1}$ is the reduced electron-exciton mass and $\Lambda$ is a momentum cut-off.
In deriving Eq.~\eqref{eq:Tmatrix}, we have assumed that the  bare exciton-electron interaction $V_\mathbf{q}$ is a constant for the relevant momenta and  expressed it in terms of the electron-exciton pair propagator in a vacuum $\Pi_v$ evaluated at the vacuum trion energy $\varepsilon_v^{T}$~\cite{Wouters2007,Carusotto2010}.

Performing the Matsubara sum in Eq.~\ref{Pairpropagtor} yields
\begin{gather} \label{eq:a1}
\Pi(k)=\sum_{i}\int \frac{d^{2}\mathbf{q}}{(2\pi)^{2}}(\mathcal{X}_{i,\mathbf{k}+\mathbf{q}}^{0})^{2}\frac{1+g(\xi^{0}_{i\mathbf{k}+\mathbf{q}})-f(\xi_{e\,-\mathbf{q}}) }{i\kappa_{q}-\xi^{0}_{i\mathbf{k}+\mathbf{q}}-\xi_{e-\mathbf{q}}},
\end{gather}
where $g(x)=(\exp x-1)^{-1}$ is the Bose-Einstein distribution. We have introduced the polariton chemical potential $\mu_P$ via $\xi^{0}_{i\mathbf k}=\varepsilon^{0}_{i\mathbf k}-\mu_P$ to account for a non-zero concentration of the polaritons, which is obviously necessary in order to explore  the interaction between them. Initially, we fix this potential for a very small but finite concentration of polaritons by employing the exciton propagator in absence of interactions $n_{L}=T\sum_{q}\mathcal{G}_{x}^{0}(q)$. Then, we self-consistently correct the chemical potential for increasing density by considering the energy of the interacting polaritons. Finally, because the Hopfield coefficients tend rapidly to their asymptotic values ($\mathcal{C}_{\mathbf{k}}^{0}\rightarrow 1$ and $\mathcal{S}_{\mathbf{k}}^{0}\rightarrow 0$) as the momentum increases, the term involving $\mathcal{S}_{\mathbf{k}}^{0}$ in $\mathcal{G}_{x}^{0}(q)$ provides a small contribution to the real part of the propagator. It can thus be neglected to an excellent approximation. 
A similar approach has been employed to describe polaron-polaritons in a Bose-Einstein condensate of polaritons~\cite{Bastarrachea-Magnani2019}.

The pole of the scattering matrix determines the energy of the trion $\varepsilon_{T\mathbf{k}}$, which differs from the bare one $\varepsilon_{T\mathbf{k}}^0$ 
due to the presence of the 2DEG and the light coupling. Pauli blocking from the 2DEG means that states with momenta below the Fermi momentum $k_F$ cannot 
contribute to forming the trion. Since the density of states is constant in 2D, we can estimate the energy shift of a zero momentum trion due to this Pauli blocking 
as  $\varepsilon_{T\mathbf{k}=0}=\varepsilon_{T\mathbf{k}=0}^0+k_F^2/2m_x+k_F^2/2m_e $.

\section{Polariton branches and coupling to the trion}
As it is defined in the main text, the polariton Green's function in frequency space is given by 
\begin{align}\label{GreensFn}
\mathcal{G}^{-1}(k)=
\begin{bmatrix}
i\omega_k-\varepsilon_{x\mathbf{k}}- \Sigma(k)& \Omega \\
\Omega & i\omega_k-\varepsilon_{c\mathbf{k}}
\end{bmatrix}.
\end{align}
In diagonal form it becomes
\begin{align}
\mathcal{G}^{-1}(k)=
\begin{bmatrix}
i\omega_k-\varepsilon_{L\mathbf{k}}& 0 \\
0 & i\omega_k-\varepsilon_{U\mathbf{k}},
\end{bmatrix}
\end{align}
where the energies of the quasi-particles are given by the self-consistent solutions of
\begin{gather} \label{eq:pols1}
\varepsilon_{L\mathbf{k}}=\frac{1}{2}\left[\varepsilon_{c\mathbf{k}}+\varepsilon_{x\mathbf{k}}+\Sigma\left(\mathbf{k},\varepsilon_{L\mathbf{k}}\right)-
\sqrt{\left[\varepsilon_{c\mathbf{k}}-\varepsilon_{x\mathbf{k}}-\Sigma\left(\mathbf{k},\varepsilon_{L\mathbf{k}}\right)\right]^{2}+4\Omega^{2}}\right], \\ 
\label{eq:pols2}
\varepsilon_{U\mathbf{k}}=\frac{1}{2}\left[\varepsilon_{c\mathbf{k}}+\varepsilon_{x\mathbf{k}}+\Sigma\left(\mathbf{k},\varepsilon_{U\mathbf{k}}\right)+
\sqrt{\left[\varepsilon_{c\mathbf{k}}-\varepsilon_{x\mathbf{k}}-\Sigma\left(\mathbf{k},\varepsilon_{U\mathbf{k}}\right)\right]^{2}+4\Omega^{2}}\right].
\end{gather}
Depending on the detuning and the Fermi energy, either   Eq.~\eqref{eq:pols1} or Eq.~\eqref{eq:pols2} has two solutions giving rise to three quasiparticle 
branches in total. 

A polariton gets damped when its energy $\varepsilon_{\mathbf p}$ plus that of an electron from the 2DEG can make a trion. Considering a zero momentum 
polariton, this gives the condition $\varepsilon_{\mathbf{p}=0}+q^2/2m_e=\varepsilon_{T\mathbf{k}=0}^0+k_F^2/2m_x+k_F^2/2m_e+q^2/2m_T$, where 
$q$ is the momentum of the electron and we have estimated the energy of a trion with momentum $q$ to be   
$\varepsilon_{T\mathbf{k}=0}^0+k_F^2/2m_x+k_F^2/2m_e+q^2/2m_T$. Taking $q=k_F$ gives the minimum energy
\begin{gather}
\varepsilon_{T}=\varepsilon_{T}^{0}+\left(m_{e}/m_{x}+m_{e}/m_{T}\right)\mu_{e}.
\end{gather}
 for which a polariton gets damped due to coupling to the 2DEG. This energy is indicated by a green horizontal line in Fig.\ 2 of the main manuscript.

%%%%%%%%%%%%%%%%%%%%%%%%%
%%%%%%%%%%%%%%%%%%%%%%%%%
%%%%%%%%%%%%%%%%%%%%%%%%%

\section{Polariton-polariton interaction for weak coupling}
\label{app:3}
The self-energy depends of the density of the polaritons via the pair propagator in Eq.~\eqref{eq:a1}. 
To make this dependence explicit, we write 
 $\Pi(k)=\Pi_{0}(k)+\delta \Pi(k)$ 
 where \begin{gather} 
\Pi_{0}(k)=\sum_{i}\int \frac{d^{2}\mathbf{q}}{(2\pi)^{2}}\mathcal{X}_{i,\mathbf{k}+\mathbf{q}}^{2}\frac{1-f(\xi_{e\,-\mathbf{q}}) }{i\kappa_{q}-\xi_{i\mathbf{k}+\mathbf{q}}-\xi_{e\,-\mathbf{q}}},
\end{gather}
is the pair propagator for zero polariton density and  
\begin{gather} 
\delta\Pi(k)=\sum_{i}\int \frac{d^{2}\mathbf{q}}{(2\pi)^{2}}\mathcal{X}^{2}_{i,\mathbf{k}+\mathbf{q}}\frac{f(\xi_{i\mathbf{k}+\mathbf{q}})}{i\kappa_{q}-\xi_{i\mathbf{k}+\mathbf{q}}-\xi_{e-\mathbf{q}}}.
\end{gather}
gives the correction due to a non-zero polariton density. 
By considering $\delta\Pi(k)$ as a small correction, the scattering matrix can be approximated as
\begin{gather} \label{eq:a2d1}
\mathcal{T}(k)=\mathcal{T}_{n_{L}=0}(k)+\mathcal{T}^{2}_{n_{L}=0}(k)\delta\Pi(k),
\end{gather}
where 
\begin{gather}
\mathcal{T}_{n_{L}=0}(k)=\mathcal{T}_{0}\left(1-\mathcal{T}_{0}\Pi_{0}(k)\right)^{-1}
\end{gather}
is the many-body scattering matrix for zero density of polaritons. Now, we substitute Eq.~\eqref{eq:a2d1} into Eq.~(3) of the main text
and neglect the scattering matrix pole contribution. This leads to

\begin{gather} \label{eq:a2d2}
\Sigma(k)=\Sigma_{n_{L}=0}(k)+
\sum_{i}\int\frac{d^{2}\mathbf{q}}{(2\pi)^{2}}\int\frac{d^{2}\mathbf{p}}{(2\pi)^{2}}
\frac{\mathcal{X}^{2}_{i,\mathbf{k}+\mathbf{q}+\mathbf{p}}\mathcal{T}_{n_{L}=0}^{2}(\mathbf{k}+\mathbf{q},i\omega_{k}+\xi_{e\mathbf{q}}) f(\xi_{e\mathbf{q}}) g(\xi_{i\mathbf{k}+\mathbf{q}+\mathbf{p}})}{i\omega_{k}+\xi_{e\mathbf{q}}-\xi_{e-\mathbf{p}}-\xi_{i\mathbf{k}+\mathbf{q}+\mathbf{p}}}\\ \nonumber
-\int\frac{d^{2}\mathbf{q}}{(2\pi)^{2}}\int_{-\infty}^{\infty}\frac{d\omega'}{\pi}\frac{f(\omega')\mbox{Im}\left[\mathcal{T}_{n_{L}=0}(\mathbf{k}+\mathbf{q},\omega'+i0^{+})+\mathcal{T}^{2}_{n_{L}=0}(\mathbf{k}+\mathbf{q},\omega'+i0^{+})\delta\Pi(\mathbf{k}+\mathbf{q},\omega'+i0^{+})\right]}{\omega'-i\omega_{k}-\xi_{e\mathbf{q}}}.
\end{gather}

Next, we use the following relations and approximate them just to keep terms of the order $\mathcal{T}^{2}_{n_{L}=0}$
\begin{gather}
\mbox{Im}\mathcal{T}_{n_{L}=0}=|| \mathcal{T}_{n_{L}=0}||^{2}\mbox{Im}\Pi_{0}\simeq \mathcal{T}^{2}_{n_{L}=0}\mbox{Im}\Pi_{0},
\end{gather}
\begin{gather}
\mbox{Im}\left(\mathcal{T}^{2}_{n_{L}=0}\delta\Pi \right)=\mathcal{T}^{2}_{n_{L}=0}\mbox{Im}\delta\Pi+\mbox{Im}\mathcal{T}^{2}_{n_{L}=0}\delta\Pi^{*}\simeq \\ \nonumber
\mathcal{T}^{2}_{n_{L}=0}\mbox{Im}\delta \Pi+\mathcal{O}(n_{L,U}^{2}).
\end{gather}
Using them, we can write Eq.~\eqref{eq:a2d2} as

\begin{gather}
\Sigma(k)\simeq\Sigma_{n_{L}=0}(k)+
\sum_{i}\int\frac{d^{2}\mathbf{q}}{(2\pi)^{2}}\int\frac{d^{2}\mathbf{p}}{(2\pi)^{2}}\mathcal{X}^{2}_{i,\mathbf{k}+\mathbf{q}+\mathbf{p}}\left\{
\frac{f(\xi_{e\mathbf{q}}) g(\xi_{i\mathbf{k}+\mathbf{q}+\mathbf{p}})\mathcal{T}_{n_{L}=0}^{2}(\mathbf{k}+\mathbf{q},i\omega_{k}+\xi_{e\mathbf{q}}) }{i\omega_{k}+\xi_{e\mathbf{q}}-\xi_{e-\mathbf{p}}-\xi_{i\mathbf{k}+\mathbf{q}+\mathbf{p}}}\right.\\ \nonumber
\left.-\frac{\left[f(\xi_{e\mathbf{-p}}+\xi_{i\mathbf{k}+\mathbf{q}+\mathbf{p}})\left(1-f(\xi_{e-\mathbf{p}})+g(\xi_{i\mathbf{k}+\mathbf{q}+\mathbf{p}})\right)\right]\mathcal{T}^{2}_{n_{L}=0}(\mathbf{k}+\mathbf{q},\xi_{e\mathbf{-p}}+\xi_{i\mathbf{k}+\mathbf{q}+\mathbf{p}}+i0^{+})}{i\omega_{k}+\xi_{e\mathbf{q}}-\xi_{e\mathbf{-p}}-\xi_{i\mathbf{k}+\mathbf{q}+\mathbf{p}}} \right\} +\mathcal{O}(n_{L,U}^{2}).
\end{gather}

By using that $n_{F}(x+y)(1-n_{F}(x)+n_{B}(y))=n_{F}(x)n_{B}(y)$, we finally get
\begin{gather}
\Sigma(k)\simeq\Sigma_{n_{L}=0}(k)+
\sum_{i}\int\frac{d^{2}\mathbf{q}}{(2\pi)^{2}}g(\xi_{i\mathbf{q}}) V_{L,i}(\mathbf{k},\omega_{k};\mathbf{q},\xi_{i\mathbf{q}}),
\end{gather}
with 
\begin{gather} 
V_{L,i}(\mathbf{k},\omega_{k};\mathbf{q},\varepsilon_{i\mathbf{q}})=\mathcal{X}^{2}_{i,\mathbf{q}}\int\frac{d^{2}\mathbf{p}}{(2\pi)^{2}}\frac{f(\xi_{e\mathbf{p}})\mathcal{T}_{n_{L}=0}^{2}(\mathbf{q}-\mathbf{p},i\omega_{\nu}+\xi_{e\mathbf{p}}) -f(\xi_{e\mathbf{k}-\mathbf{q}+\mathbf{p}})\mathcal{T}^{2}_{n_{L}=0}(\mathbf{q}-\mathbf{p},\xi_{e\mathbf{k}-\mathbf{q}+\mathbf{p}}+\xi_{i\mathbf{q}}+i0^{+})}{i\omega_{\nu}-\xi_{i\mathbf{q}}+\xi_{e\mathbf{p}}-\xi_{e\mathbf{k}-\mathbf{q}+\mathbf{p}}},
\end{gather}
By taking the functional derivative with respect to distribution of polaritons in the $j$-polariton branch on-shell we obtain
\begin{gather} \label{eq:4d4}
\mathsf f_{L\mathbf k,j\mathbf k'}=\mathcal{Z}_{L\mathbf{k}}\frac{\delta\varepsilon_{L\mathbf{k}}}{ \delta n_{j{\mathbf{k}'}}}=\mathcal{Z}_{L\mathbf{k}}\mathcal{C}_{\mathbf{k}}^{2}\frac{\delta\Sigma(\mathbf{k},\varepsilon_{L\mathbf{k}})}{ \delta n_{j{\mathbf{k}'}}}=\mathcal{Z}_{L\mathbf{k}}\mathcal{C}_{\mathbf{k}}^{2}\mathcal{X}_{j\mathbf{k}'}^{2}V_{L,j}(\mathbf{k},\varepsilon_{L\mathbf{k}};\mathbf{k}',\varepsilon_{j\mathbf{k}'}).
\end{gather}
When we consider energies far from the trion energy, we can approximate the scattering matrix as a momentum-independent constant $\mathcal{T}_{n_{L}=0}$. 
In this case, Eq.~\eqref{eq:4d4} reduces to Eq.~(6) from the main text. 

%%%%%%%%%%%%%%%%%%%%%%%%%%%%%%%%%%%%%%%%%%%%%%%%%%

\end{document}